\documentstyle[12pt,epsfig]{article}
\begin{document}
\setlength{\baselineskip}{0.75cm}
\setlength{\parskip}{0.45cm}
\begin{titlepage}
\begin{flushright}
\begin{tabular}{l}
CERN-TH/97-132 \\
hep-ph/9706511
\end{tabular}
\end{flushright}
\vspace*{1.2cm}
\begin{center}
\Large{
{\bf Next-to-leading Order Evolution\\} 
\vspace*{0.5cm}
{\bf of Transversity Distributions\\}
\vspace*{0.5cm}
{\bf and Soffer's Inequality}}
\vskip 3cm
{\large Werner Vogelsang}  \\
\vspace*{1cm}
\normalsize
Theory Division, CERN,\\
CH-1211 Geneva 23, Switzerland \\
\end{center}
\vskip 1.8cm
\begin{center}
{\bf Abstract} \\
\end{center}
We present a calculation of the two-loop splitting functions for the 
evolution of the twist-2 'transversity' parton densities of transversely
polarized nucleons. We study the implications of our results for 
Soffer's inequality for the case of valence quark densities. 
CERN-TH/97-132 \\
June 1997 
\end{titlepage}
\section{Introduction}
The past few years have seen much progress in our knowledge about the
nucleon's spin structure thanks to the experimental study of the spin 
asymmetries $A_1^N (x,Q^2)$ ($N=p,n,d$) in deep-inelastic scattering (DIS) 
with longitudinally polarized lepton beams and nucleon targets. 
Such experiments yield information on the 'longitudinally polarized'
(or 'helicity-weighted') parton distributions of the nucleon, 
which will be denoted as
\begin{equation} \label{partonlong}
\Delta_L f(x,Q^2) \equiv f^{\rightarrow} (x,Q^2) - f^{\leftarrow} (x,Q^2) \; ,
\end{equation}
where $f^{\rightarrow} (x,Q^2)$ ($f^{\leftarrow} (x,Q^2)$) is the 
probability at scale $Q$ of finding a
parton-type $f$ ($f=q,\bar{q},g$) in a longitudinally polarized nucleon 
(for definiteness, say, a proton) with its spin aligned (anti-aligned) with 
the proton spin and carrying the fraction $x$ of the proton's momentum.
Taking the sum instead of the difference in Eq.~(\ref{partonlong}) one
obtains the usual unpolarized parton distributions of the proton,
\begin{equation} \label{partonunp}
f(x,Q^2) \equiv f^{\rightarrow} (x,Q^2) + f^{\leftarrow} (x,Q^2) \; .
\end{equation}

Apart from $f(x,Q^2)$ and $\Delta_L f(x,Q^2)$, a third type of parton 
density can be defined which describes the properties of {\em transversely} 
polarized nucleons \cite{rs,am,jaffe,cortes}:
\begin{equation} \label{partontrans}
\Delta_T f(x,Q^2) \equiv f^{\uparrow} (x,Q^2) - f^{\downarrow} (x,Q^2) \; ,
\end{equation}
where now $f^{\uparrow} (x,Q^2)$ ($f^{\downarrow} (x,Q^2)$) stands for
the probability of finding a parton-type $f$ in a {\em transversely} 
polarized proton with its spin parallel (antiparallel) to the proton spin.
These parton densities are usually referred to as 'transversity' 
distributions. They complete the twist-2 sector of nucleonic parton
distributions and are therefore as interesting in principle as $f(x,Q^2)$ and 
$\Delta_L f(x,Q^2)$. Unlike the case of the latter two, there is no 
gluonic transversity distribution at leading twist \cite{ji,soffer1}. 

Unfortunately, nothing is known as yet experimentally about the transversity
distributions. As was shown in \cite{am,jaffe,cortes}, it is not 
possible to measure them in inclusive deep-inelastic scattering because 
they are chirally odd. It is widely accepted that Drell-Yan lepton pair 
production in collisions of transversely polarized protons offers the best 
possibility to get access to the $\Delta_T f$ 
\cite{rs,am,jaffe,cortes,sivers,alex,cont1,dyrefs}, and the RHIC spin physics 
programme comprises experiments of this kind \cite{rhic}.    

There being a shortage of experimental information on the transversity 
distributions, there has been quite some effort on the theoretical side, 
aiming at the question of whether the $\Delta_L f$ and $\Delta_T f$ are 
entirely independent of each other. This is not expected really because
one clearly has  
\begin{equation} \label{equal}
f^{\rightarrow} (x,Q^2) + f^{\leftarrow} (x,Q^2)  = 
f^{\uparrow} (x,Q^2) + f^{\downarrow} (x,Q^2) \; .
\end{equation}
In fact, Soffer has derived an interesting non-trivial inequality 
\cite{soffer2,sivers}, 
\begin{equation} \label{si}
|\Delta_T f| \leq \frac{1}{2} \left( f + \Delta_L f \right) 
\equiv f^{\rightarrow} \; ,
\end{equation}
which is valid separately for each quark flavour. 

Soffer's inequality can be derived in the context of a parton model
\cite{soffer2,sivers,gold}. The question immediately rises of whether
the inequality is maintained when QCD is applied \cite{gold,cont2,bar}. 
There are two potential sources of concern here. Firstly, as indicated
in Eqs.~(\ref{partonlong})-(\ref{partontrans}), the parton distributions
evolve with $Q^2$. Since the corresponding evolution kernels at leading 
order (LO) \cite{ap,am} turn out to be different for the $f(x,Q^2)$, 
$\Delta_L f(x,Q^2)$, $\Delta_T f(x,Q^2)$, it is not a priori clear that
the inequality is preserved by QCD evolution. It turns out \cite{bar},
however, that this is the case. The crucial point here is that the 
difference of the involved flavour non-singlet LO Altarelli-Parisi evolution 
kernels is always positive, 
\begin{equation} \label{lodiff}
\frac{1}{2} \left( P_{qq}^{(0)} (x) + \Delta_L P_{qq}^{(0)} (x) \right) -
\Delta_T P_{qq}^{(0)} (x) = C_F (1-x) \geq 0 \; ,
\end{equation}
where $C_F=4/3$. In Eq.~(\ref{lodiff}) $P_{qq}^{(0)}$, 
$\Delta_L P_{qq}^{(0)}$ are the unpolarized and longitudinally polarized 
LO quark-to-quark splitting functions, respectively, which have been 
calculated in \cite{ap} and are equal because of helicity conservation. 
$\Delta_T P_{qq}^{(0)}$ is their transversely polarized counterpart of 
\cite{am}. From Eq.(\ref{lodiff}) it follows that if one assumes validity of 
Soffer's inequality (\ref{si}) at some input scale $Q_0$, the inequality 
will still be fulfilled at any higher scale, at least in the context of 
LO evolution. Strictly speaking, this statement only follows 
for flavour {\em non-singlet}, in particular valence, quark densities, 
which is due to the lack of gluons in the case of 
transversity distributions. However, as was shown in \cite{bar}, 
the picture does not change when the full LO singlet evolution is
taken into account for the $f(x,Q^2)$, $\Delta_L f(x,Q^2)$ on the 
right-hand side of (\ref{si}). 

Secondly, there is the question of what role higher order QCD corrections
play in Soffer's inequality. These enter in two ways, namely via the 
corrections to the splitting functions governing the evolution of the 
parton distributions, {\em and} as corrections to the cross sections for the 
processes which are used to extract information on the parton distributions 
from experiment in order to see whether they satisfy Eq.~(\ref{si}). As 
is well-known, the split-up of the higher order corrections into corrections
to the splitting functions and to the cross sections is not unique but depends 
on the factorization scheme adopted, which in turn implies that the parton
densities themselves are scheme-dependent beyond LO. Clearly, one is free 
to choose the factorization schemes independently for the $f(x,Q^2)$, 
$\Delta_L f(x,Q^2)$ and $\Delta_T f(x,Q^2)$, and it will in principle always 
be possible to violate Eq.~(\ref{si}) by an arbitrarily large amount by 
adopting certain 'sufficiently incompatible' schemes for the three densities. 
One therefore has to ask in which factorization scheme(s) it makes sense to 
try to check Eq.~(\ref{si}) 
beyond LO. In other words, what should be the meaning of Eq.~(\ref{si}) in 
higher orders? This question has also been addressed in \cite{cont2} and will 
be one key issue of this paper. We will provide all ingredients for the 
next-to-leading order (NLO) framework of transversity distributions, which 
will serve to enable a check of Soffer's inequality to NLO.

As was pointed out earlier, the Drell-Yan process with transversely polarized
hadrons appears to be the most likely candidate for measurements of the
transversity distributions. Much theoretical work has therefore focused on 
this process \cite{rs,am,jaffe,cortes,sivers,alex,cont1,dyrefs}. 
In particular, the NLO corrections to the transversely polarized Drell-Yan 
have been calculated in \cite{alex} and \cite{cont1,basim}, within 
different factorization schemes. The result of the latter calculation 
\cite{cont1,basim} was used in \cite{cont2} for a first 
assessment of the impact of NLO corrections on Soffer's inequality. 
However, for future NLO analyses of (hopefully forthcoming) experimental
data, knowledge of the NLO corrections to the Drell-Yan cross section
alone cannot be sufficient as data will necessarily be taken at various 
different values of the invariant mass of the Drell-Yan muon pair
which sets the hard scale for the process. According to the previous paragraph,
the NLO corrections to the $Q^2$-evolution of the transversity distributions
also have to be known in order to perform a complete and consistent NLO
calculation. The main result of this paper will therefore be the presentation 
of the NLO evolution kernels for the transversity distributions -- only when 
these are available can Soffer's inequality be meaningfully studied
beyond LO.
The method we will use to calculate the NLO transversity splitting functions  
will be the one used by Curci, Furmanski and Petronzio in the unpolarized 
case \cite{cfp,fp} (see also \cite{ev}) and which was also utilized 
recently \cite{wv} to obtain the NLO evolution kernels for the longitudinally 
polarized parton distributions $\Delta_L f$.    

The remainder of this paper is organized as follows. In section 2 we will
outline the framework for our calculations. Also in our case the transversely
polarized Drell-Yan process will be the starting point. Section 3 will
give details of the calculation of the NLO transversity splitting functions.
Section 4 will present the results and focus on the validity of Soffer's 
inequality beyond LO. Finally, we will draw our conclusions in section 5.
\section{Framework}
We begin by defining the spin-dependent cross section for dimuon 
production by protons with transverse polarization along, say, the 
$x$-axis:
\begin{equation} \label{wqdef}
\frac{\tau d\Delta_T \sigma}{d \tau d\phi} \equiv
\frac{1}{2} \left( \frac{\tau d\sigma^{\uparrow \uparrow}}{d 
\tau d\phi} - \frac{\tau d\sigma^{\uparrow \downarrow}}{d \tau d\phi}
\right) \; ,
\end{equation}
where the superscript $\uparrow \uparrow$ ($\uparrow \downarrow$)
denotes parallel (antiparallel) setting of the transverse spins
of the incoming protons. Furthermore, $\tau \equiv M^2/S$, where 
$M$ is the invariant mass of the muon pair and $\sqrt{S}$ the
centre-of-mass energy of the hadronic collision; $\phi$ is the azimuthal 
angle of one of the muons and is counted relative to the axis defined 
by the transverse polarizations. For the discussion to follow in this section
it is convenient to temporarily go to Mellin moment space where we can
perform the $Q^2$-evolutions more explicitly. The Mellin moments of the 
cross section are defined by
\begin{equation} \label{mellin}
\frac{d\Delta_T \sigma^n}{d\phi} \equiv \int_0^1 d\tau \tau^{n-1}
\frac{\tau d\Delta_T \sigma}{d \tau d\phi} \; .
\end{equation}
Including the NLO corrections to this cross section (as calculated in 
a certain factorization scheme such as the $\overline{{\mbox{\rm MS}}}$ 
scheme), one has the generic expression\footnote{Here we only consider photon 
exchange. The contributions from $Z^0$ exchange and $\gamma$-$Z^0$ 
interference can be straightforwardly included by making use of Eq.~(3) of 
\cite{cont1}.} 
\cite{alex,cont1}
\begin{equation} \label{nlogen}
\frac{d\Delta_T \sigma^n}{d\phi} = \frac{\alpha_{em}^2}{9S} \cos (2 \phi)
\Delta_T H^n (M^2) \left( 1 + \frac{\alpha_s (M^2)}{2\pi} \Delta_T C_q^{
{\small {\rm DY}},n} 
\right) \; ,
\end{equation}
where 
\begin{equation} \label{hdef}
\Delta_T H^n (Q^2) \equiv \sum_q e_q^2 \left[ \Delta_T q_{1}^n (Q^2) 
\Delta_T \bar{q}_{2}^n (Q^2) + \Delta_T \bar{q}_{1}^n (Q^2)
\Delta_T q_{2}^n (Q^2) \right] \; .
\end{equation}
Here we have restored for the moment the possibility of the scattering of
two {\em different} hadrons labelled '1','2', even though anything apart from
$pp$ scattering will presumably not be realistic experimentally. We will keep
this notation when dealing with the function $\Delta_T H$, but otherwise 
suppress the indices '1','2' in the following. The Mellin moments of the 
quark distributions $\Delta_T q$ and the ${\cal O}(\alpha_s)$ corrections 
$\Delta_T C_q^{{\small {\rm DY}}}$ are defined in analogy with 
Eq.~(\ref{mellin}), e.g. $\Delta_T q^n (Q^2) \equiv \int_0^1 dx x^{n-1} 
\Delta_T q(x,Q^2)$. 
 
When including the ${\cal O}(\alpha_s)$ corrections in (\ref{nlogen}),
it is crucial that the parton distributions are evolved to NLO accuracy as 
well. Since there are no gluons involved in the case of the transversity 
distributions, their evolution equations reduce to simple non-singlet type 
equations. Introducing 
\begin{equation}
\Delta_T q_{\pm}^{n} \equiv \Delta_T q^n \pm \Delta_T \bar{q}^n \; ,
\end{equation}
one has
\begin{equation} \label{hdef1}
\Delta_T H^n (Q^2) = \frac{1}{2} \sum_q e_q^2 \Bigg[ \Delta_T q_{+,1}^n (Q^2)
\Delta_T q_{+,2}^n (Q^2) - \Delta_T q_{-,1}^n (Q^2) \Delta_T q_{-,2}^n (Q^2) 
\Bigg]
\end{equation}
and the evolution equations (see, e.g., \cite{ev})
\begin{eqnarray} 
\frac{d}{d\ln Q^2} \Delta_T q_-^n (Q^2) &=& \Delta_T P_{qq,-}^n 
(\alpha_s (Q^2)) \Delta_T q_-^n (Q^2) \label{evol2} \; , \\
\frac{d}{d\ln Q^2} ( \Delta_T q_+^n - \Delta_T q'{}_+^{n} ) (Q^2) &=& 
\Delta_T P_{qq,+}^n (\alpha_s (Q^2)) ( \Delta_T q_+^n - 
\Delta_T q'{}_+^{n} ) (Q^2) \; , \label{evol1} \\
\frac{d}{d\ln Q^2} \Delta_T \Sigma^n (Q^2) &=& \Delta_T P_{\Sigma\Sigma}^n 
(\alpha_s (Q^2)) \Delta_T \Sigma^n (Q^2) \; , \label{singev} 
\end{eqnarray}
where $q,q'$ are different quark flavours and $\Delta_T \Sigma^n \equiv 
\sum_q \Delta_T q_+^n$. Note that the first
moment ($n=1$) in Eq.~(\ref{evol2}) corresponds to the evolution
of the nucleon's tensor charge \cite{jaffe,soffer3}. 
The splitting functions $\Delta_T P_{qq,\pm} (\alpha_s (Q^2))$, 
$\Delta_T P_{\Sigma\Sigma}$ are taken to have the following perturbative 
expansion:
\begin{equation} \label{expan}
\Delta_T P_{ii}^n (\alpha_s) = \left( \frac{\alpha_s}{2\pi} \right) \Delta_T 
P_{qq}^{(0),n} + \left( \frac{\alpha_s}{2\pi} \right)^2 \Delta_T
P_{ii}^{(1),n} + \ldots  \; , 
\end{equation}
$\{ ii \} = \{ qq,\pm \}, \{\Sigma \Sigma \}$. As indicated, 
$\Delta_T P_{qq,+}$, $\Delta_T P_{qq,-}$ and $\Delta_T
P_{\Sigma\Sigma}$ are all equal at LO. It is convenient to write \cite{ev}
\begin{eqnarray} \label{pm}
\Delta_T P_{qq,\pm}^{(1),n} &\equiv& \Delta_T P_{qq}^{(1),n} \pm \Delta_T 
P_{q\bar{q}}^{(1),n} \; , \\
\Delta_T P_{\Sigma\Sigma}^{(1),n} &\equiv& \Delta_T P_{qq,+}^{(1),n} + 
\Delta_T P_{qq,PS}^{(1),n} \; .
\end{eqnarray}
It will actually turn out that $\Delta_T P_{qq,PS}^{(1),n} \equiv 0$, so
that Eqs.~(\ref{evol1}),(\ref{singev}) can be replaced by
\begin{equation} \label{evol3}
\frac{d}{d\ln Q^2} \Delta_T q_+^n (Q^2) = \Delta_T P_{qq,+}^n 
(\alpha_s (Q^2)) \Delta_T q_+^n (Q^2) \; .
\end{equation}
Needless to say that the
NLO splitting functions have to be known in the same factorization scheme
as the corrections to the cross section,  $\Delta_T C_q^{{\small {\rm DY}}}$.

The solution to Eqs.~(\ref{evol2}),(\ref{evol3}) is well-known (see, e.g.,
\cite{gr}) and reads
\begin{eqnarray} \label{evsol}
\nonumber
\Delta_T q_{\pm}^n (Q^2) &=& \left( 1+\frac{\alpha_s (Q_0^2)-
\alpha_s (Q^2)}{\pi\beta_0}\!
\left[ \Delta_T P_{qq,\pm}^{(1),n}-\frac{\beta_1}{2\beta_0} \Delta_T 
P_{qq}^{(0),n} \right] \right)\\
&&\times
\left( \frac{\alpha_s (Q^2)}{\alpha_s (Q_0^2)}\right)^{-2\Delta_T 
P_{qq}^{(0),n}/ \beta_0}\! \Delta_T q_{\pm}^n (Q_0^2)
\end{eqnarray}
with the input distributions $\Delta_T q_{\pm}^n (Q_0^2)$ at the input scale 
$Q_0$. Here and in all previous NLO equations one has to use the two-loop 
expression for the strong coupling,
\begin{equation}
\frac{\alpha_s(Q^2)}{4\pi} \simeq \frac{1}{\beta_0
\ln Q^2/\Lambda_{\overline{\rm{MS}}}^2}-
\frac{\beta_1}{\beta_0^3} \frac{\ln\ln Q^2/\Lambda_{\overline{\rm{MS}}}^2}
{\left( \ln Q^2/\Lambda_{\overline{\rm{MS}}}^2\right)^2}
\end{equation}
with the QCD scale parameter $\Lambda_{\overline{\rm{MS}}}$ and $\beta_0 = 
11-2f/3$, $\beta_1=102-38f/3$, $f$ being the number of active flavours.
We note that all LO expressions are entailed in the above equations by
setting the NLO quantities $\Delta_T C_q^{{\small {\rm DY}}}$, 
$\Delta_T P_{\ldots}^{(1)}$, $\beta_1$, to zero.

Formulas exactly analogous to the above hold for the unpolarized and the 
longitudinally polarized cases. The only complication here is that there
are obviously contributions by gluons to the Drell-Yan cross section and to 
the evolution of the quark singlet combination $(\Delta_L) \Sigma^n$. Let us 
assume for the moment that we are able to eliminate these experimentally by 
taking suitable non-singlet combinations of Drell-Yan cross sections for
various configurations of scattering hadrons. In particular, when considering
the difference of the cross sections for $p\bar{p}$ and $pp$ scattering,
the function $\Delta_T H^n$ reduces to pure valence,
\begin{equation} \label{hdef2}
\Delta_T H^n (Q^2) = \sum_q e_q^2 \Delta_T q_{-}^n (Q^2) \Delta_T 
q_{-}^n (Q^2) \; ,
\end{equation}
where all parton distributions refer to the proton, and with similar
expressions for the unpolarized and longitudinally polarized cases. 
We then only need to consider Eqs.~(\ref{nlogen}),(\ref{evol2}) which 
exactly carry over to the unpolarized case (by omitting all $\Delta_T$ 
and replacing $\cos (2\phi)$ by $2$) or to the longitudinally 
polarized case (by replacing $\Delta_T$ by $\Delta_L$ and $\cos (2\phi)$
by $-2$), respectively. 

We now adopt the Drell-Yan process as the process defining the (NLO) parton 
distributions and follow \cite{cont2} to argue that (if at all) Soffer's 
inequality for the valence densities should be valid beyond LO {\em if} 
we choose a factorization scheme in which the non-singlet Drell-Yan 
cross sections for the unpolarized, the longitudinally polarized and 
the transversely polarized cases all individually maintain their respective 
LO forms. This appears reasonable since the parton distributions are then 
directly related to a physical (observable) quantity. For example, in case 
of the cross section for transverse polarization it means 
\begin{equation} \label{nlogen1}
\frac{d\Delta_T \sigma^n}{d\phi} = \frac{\alpha_{em}^2}{9S} \cos (2 \phi)
\Delta_T \tilde{H}^n (M^2) \; ,
\end{equation}
where $\Delta_T \tilde{H}^n (Q^2)$ is defined as in (\ref{hdef}),(\ref{hdef2}) 
but in terms of the NLO parton densities $\Delta_T \tilde{q}^n (Q^2)$ 
in the new scheme. Inserting (\ref{evsol}) into (\ref{nlogen}), expanding in 
$\alpha_s$ and equating with (\ref{nlogen1}) we find 
\begin{equation} \label{evsol1}
\Delta_T \tilde{q}_{-}^n (Q^2) = \left( 1+\frac{\alpha_s (Q_0^2)-
\alpha_s (Q^2)}{\pi\beta_0}\! \Delta_T {\cal E}_{-}^n  \right)
\left( \frac{\alpha_s (Q^2)}{\alpha_s (Q_0^2)}\right)^{-2\Delta_T 
P_{qq}^{(0),n}/ \beta_0}\! \Delta_T \tilde{q}_{-}^n (Q_0^2) 
\end{equation}
in the new scheme, where 
\begin{equation} \label{edef}
\Delta_T {\cal E}_{-}^n \equiv \Delta_T P_{qq,-}^{(1),n}-\frac{\beta_0}{4}
\Delta_T C_q^{{\small {\rm DY}},n} - \frac{\beta_1}{2\beta_0} 
\Delta_T P_{qq}^{(0),n} 
\end{equation}
is a combination of ($\overline{{\mbox{\rm MS}}}$) NLO quantities which is 
manifestly independent of the factorization scheme due to the fact that the 
Drell-Yan cross section in Eq.~(\ref{nlogen1}) is a physical quantity.
As a necessary condition for Eq.~(\ref{si}) to hold for the valence densities, 
its moments (for real $n>0$) must also satisfy Soffer's inequality. Thus
\begin{eqnarray} \label{nloeq}
\nonumber
\frac{2 |\Delta_T \tilde{q}_{-}^n (Q^2)|}{\tilde{q}_{-}^n (Q^2) +
\Delta_L \tilde{q}_{-}^n (Q^2)} &=& \frac{1+( \alpha_s (Q_0^2)-
\alpha_s (Q^2)) \Delta_T {\cal E}_{-}^n/(\pi\beta_0)}
{1+( \alpha_s (Q_0^2)-\alpha_s (Q^2)) \left[ {\cal E}_{+}^n -
\frac{2 \tilde{q}_{-}^n (Q_0^2)}{\tilde{q}_{-}^n (Q_0^2) +
\Delta_L \tilde{q}_{-}^n (Q_0^2)} P_{q\bar{q}}^{(1),n} \right] 
/(\pi\beta_0)} \\ 
&&\times 
\left( \frac{\alpha_s (Q^2)}{\alpha_s (Q_0^2)}\right)^{-2 (\Delta_T 
P_{qq}^{(0),n}-P_{qq}^{(0),n}) / \beta_0}\! 
\frac{2 |\Delta_T \tilde{q}_{-}^n (Q_0^2)|}{\tilde{q}_{-}^n (Q_0^2) +
\Delta_L \tilde{q}_{-}^n (Q_0^2)}
\end{eqnarray}
has to be smaller than unity in the scheme we have defined. Here we have used
for the evolution of $\tilde{q}_{-}^n (Q^2) + \Delta_L \tilde{q}_{-}^n (Q^2)$
that \cite{ap} $\Delta_L P_{qq}^{(0),n} \equiv P_{qq}^{(0),n}$ and
\cite{rat,weber} $\Delta_L C_q^{{\small {\rm DY}},n} \equiv 
C_q^{{\small {\rm DY}},n}$ as well as \cite{svw,grsv,wv} 
$\Delta_L P_{qq,\pm}^{(1),n} \equiv P_{qq,\mp}^{(1),n}$
where $P_{qq,\pm}^{(1),n} = P_{qq}^{(1),n} \pm P_{q\bar{q}}^{(1),n}$ 
is defined in analogy with (\ref{pm}).
Clearly, the right-hand side in (\ref{nloeq}) will in general only be 
smaller than unity if this is the case for the input densities at $Q=Q_0$, 
$2 |\Delta_T \tilde{q}_-^n (Q_0^2)|/(\tilde{q}_-^n (Q_0^2) + \Delta_L 
\tilde{q}_-^n (Q_0^2)) \leq 1$. This, of course, is impossible to 
prove or disprove within perturbative QCD. What one can do (and what is the
purpose of this paper) is to assume validity of Soffer's inequality for the 
input and see whether the inequality is preserved when going to
higher $Q^2$. From (\ref{lodiff}) we know that the exponent $-2 (\Delta_T 
P_{qq}^{(0),n}-P_{qq}^{(0),n}) / \beta_0$ is positive, which demonstrates
preservation of Soffer's inequality at the LO level as discussed in
\cite{bar}. Even more, one finds that for any reasonably small 
$\alpha_s (Q_0^2)$ 
\begin{equation}
\left[ 1-\frac{\alpha_s (Q_0^2)-\alpha_s (Q^2)}{\pi\beta_0}\! \; \; 
\frac{\beta_1}{2\beta_0} \left( \Delta_T P_{qq}^{(0),n}-P_{qq}^{(0),n}
\right) \right] \left( \frac{\alpha_s (Q^2)}{\alpha_s 
(Q_0^2)}\right)^{-2 (\Delta_T P_{qq}^{(0),n}-P_{qq}^{(0),n}) / \beta_0}
\end{equation}
remains smaller than unity. It also turns out that the unpolarized 
$P_{q\bar{q}}^{(1),n}$ is always negative, so that the corresponding term 
in the denominator of Eq.~(\ref{nloeq}) will always help {\em preserve} 
Soffer's inequality. What still is to be shown is that the remaining terms in 
(\ref{nloeq}) do not cause any problem if (in the 'worst' case)
$2|\Delta_T \tilde{q}_{-}^n (Q_0^2)|/(\tilde{q}_{-}^n (Q_0^2) +
\Delta_L \tilde{q}_{-}^n (Q_0^2))=1$. Expanding in $\alpha_s$ and taking into 
account the above observations one immediately finds that in this case
\begin{equation} \label{master}
\left( P_{qq}^{(1),n}-\Delta_T P_{qq,-}^{(1),n} \right) - 
\frac{\beta_0}{4} \left( C_q^{{\small {\rm DY}},n} - 
\Delta_T C_q^{{\small {\rm DY}},n} \right) 
\end{equation} 
has to be positive to guarantee validity of Soffer's inequality for valence
densities at NLO. Having worked out this condition in Mellin-$n$ space for 
convenience, we will now return to Bj\o rken-$x$ space.
In the following we will calculate the splitting function $\Delta_T 
P_{qq,-}^{(1)} (x)$, to see whether the $x$-space counterpart of the 
expression in (\ref{master}),
\begin{equation} \label{masterx}
\left( P_{qq}^{(1)}(x)-\Delta_T P_{qq,-}^{(1)}(x) \right) - 
\frac{\beta_0}{4} \left( C_q^{{\small {\rm DY}}} (x) - \Delta_T 
C_q^{{\small {\rm DY}}} (x) \right) \; ,
\end{equation} 
is indeed positive.

We emphasize that our considerations in the above equations only
apply to the valence densities. We will also provide the splitting
function $\Delta_T P_{qq,+}^{(1)}$ which is needed for the NLO
evolution of the combination $\Delta_T q_+$ in (\ref{evol3}). To examine, 
however, the validity of Soffer's inequality for the non-valence quark 
densities would involve taking into account the singlet evolution of the 
unpolarized and longitudinally polarized parton distributions. This can only 
be done within a detailed numerical study, which is beyond the scope of this 
paper. 
\section{Calculation of the NLO transversity splitting functions}
\newcommand{\slsh}{\rlap{$\;\!\!\not$}}     
\def\Proj{\Delta_T {\cal P}}
Our calculation of the NLO splitting functions for the transversity
distributions closely follows the ones performed in the unpolarized 
\cite{cfp,fp,ev} and longitudinally polarized \cite{wv} cases.
We only briefly outline the method and its application to the transversity
case here. A thorough overview of the technique can be found in 
\cite{cfp,ev}.  

The method we will use is set up in Bj\o rken-$x$ space. It is based 
on the factorization properties of mass 
singularities in the light-like axial gauge, specified by introducing a 
light-like vector $n$ ($n^2=0$) with $n \cdot A=0$, $A$ being the gluon field.
The general strategy then consists of first expanding the squared 
matrix element $\Delta_T M$ for the scattering of two (transversely polarized)
(anti)quarks into two ladders of two-particle irreducible (2PI) 
kernels \cite{egmpr}. The crucial point is that in the light-cone 
gauge the 2PI kernels are finite before the integrations over the 
sides of the ladders are performed. Collinear singularities therefore 
appear only when integrating over the lines connecting the rungs of the 
ladders \cite{egmpr}. This allows for systematically projecting out 
the singularities by introducing a projector onto (transversely polarized) 
physical states, $\Proj$. More precisely, $\Proj$ decouples the product 
$\Delta_T (AB)$ of two successive 2PI kernels by projecting onto  
states of definite transverse polarization of the particle connecting the 
kernels and by setting this particle on-shell in kernel $A$. Transverse
polarization for a quark entering a kernel is obtained by using
\begin{equation} \label{g51}
u(p,s) \bar{u} (p,s) = - \slsh{p} \slsh{s} \gamma_5 \; ,
\end{equation}
for its spinor $u(p,s)$. Here $p$ is the quark's momentum and $s$ its 
transverse spin vector satisfying $s^2=-1$, $s \cdot p =0$, $s \cdot n=0$, 
where the latter equality is a consequence of using $n$ to define the  
longitudinal direction, 
\begin{equation}
n \cdot p \equiv pn \neq 0, \;\; n \cdot t=p \cdot t=0 \; , 
\end{equation}
$t$ being any transverse vector. The part of $\Proj$ that is relevant for 
the Dirac algebra is then found to be  
\begin{equation} \label{g52}
\Proj \sim \frac{1}{4 n \cdot k} \slsh{n} \slsh{s} \gamma_5 \; ,
\end{equation}
where $k$ is the momentum of the particle emerging from the top of the kernel.
As already mentioned several times, no gluonic projection operators
are to be introduced in the transversity case. 
By means of $\Proj$, $\Delta_T M$ can be written in the factorized form 
\begin{equation}
\Delta_T M = \Delta_T \hat{\sigma} \otimes \Delta_T \Gamma_1 \otimes 
\Delta_T \Gamma_2 \;,
\end{equation}
where $\Delta_T \hat{\sigma}$ is interpreted as the (finite) short-distance
(Drell-Yan) subprocess cross section, whereas the factors $\Delta_T 
\Gamma$ (one for each quark) contain all (and only) mass singularities and 
are process-{\em independent}. The explicit expression for $\Delta_T \Gamma$
in terms of the 2PI kernels can be easily derived from \cite{cfp,ev}.
Working in dimensional regularization 
($d=4-2 \epsilon$ dimensions) in the $\overline{\mbox{MS}}$ scheme it 
can be shown \cite{cfp} that the residue of the $1/\epsilon$ pole of 
$\Delta_T \Gamma$ corresponds to the evolution kernels we are looking for:
\begin{equation}
\Delta_T \Gamma_{ii} \left( \hspace*{-0.08cm} 
x,\alpha_s,\frac{1}{\epsilon} \right) = \delta (1-x) - 
\frac{1}{\epsilon} \Bigg(\frac{\alpha_s}{2\pi} \Delta_T P_{qq}^{(0)}(x)
+\frac{1}{2} \left( \frac{\alpha_s}{2\pi} \right)^2 \hspace*{-0.08cm} 
\Delta_T P_{ii}^{(1)} (x) + \ldots \Bigg) + O \left(\frac{1}{\epsilon^2} 
\right) \: ,
\end{equation}
where we have restored the subscripts $\{ ii \} = \{ qq,\pm \}, \{\Sigma
\Sigma \}$ that distinguish between the various NLO quark-to-quark splitting
functions. 

The Feynman graphs contributing to $\Delta_T \Gamma_{ii}$ at NLO are
obviously the same as in the unpolarized and the longitudinally polarized 
calculations \cite{cfp,ev,wv} and need not be repeated here. Let us
instead collect a few important points of the calculation\footnote{We use the 
program {\sc Tracer} of \cite{trac} for calculating the Dirac traces and 
performing contractions.}:
\begin{itemize}
\item As mentioned above, the calculation relies on the use of the 
light-cone gauge. Following \cite{cfp,ev,wv}, we use the principal value
prescription to regularize the spurious singularities resulting from the  
gauge propagator (see \cite{ev} for details). 
\item From Eqs.~(\ref{g51}),(\ref{g52}) we see that the Dirac matrix
$\gamma_5$ enters the calculation via the projectors on physical 
states of transverse polarization of the quarks. In general, this can 
lead to complications when dimensional regularization is used. It turns
out, however, that no problems related to $\gamma_5$ occur in our 
calculation due to the fact that all Dirac traces that we need contain 
{\em two} $\gamma_5$ matrices, one coming from the projector $\Proj$ in 
(\ref{g52}), the other from (\ref{g51}). In this case, it is safe to use
a fully anticommuting $\gamma_5$, which effectively removes the $\gamma_5$
from the traces via $\gamma_5^2=1$. The only case where this does not work
is the splitting function $\Delta_T P_{qq,PS}^{(1)}$ for which there are two
separate quark lines containing one $\gamma_5$ each. However, it is 
immediately obvious that for both Dirac traces the $\gamma_5$ is accompanied 
by an odd number of other Dirac matrices. So the traces vanish irrespective 
of the prescription for $\gamma_5$. As a result, the splitting function 
$\Delta_T P_{qq,PS}^{(1)}$ vanishes identically.
\item The calculation is technically slightly more involved than in the
unpolarized or the longitudinally polarized cases, owing to the appearance
of scalar products of momenta with the transverse spin vector
in the squared matrix elements. These scalar products, which are
obviously always quadratic in $s$, introduce extra transverse
degrees of freedom that need to be integrated out. In LO, this is a
rather straightforward task. The only scalar product of this kind that
occurs is $(k\cdot s)^2$, which after integration can be effectively
replaced by
\begin{equation} \label{repl1}
(k \cdot s)^2 \rightarrow -\frac{1}{2 (1-\epsilon)} k^2 (1-x) \; .
\end{equation}
This term is actually multiplied by $\epsilon$ in the LO squared matrix 
element. As a result, it only contributes to the LO transversity splitting 
function in $d=4-2 \epsilon$ dimensions, $\Delta_T P_{qq}^{(0),d=4-2 \epsilon} 
(x)$, but not to its {\em four}-dimensional counterpart. In the NLO 
calculation, however, the $d$-dimensional LO result $\Delta_T 
P_{qq}^{(0),d=4-2 \epsilon} (x)$ is needed. It turns out that the 
contribution from the term $\sim \epsilon (k \cdot s)^2$ is cancelled by 
other terms $\sim \epsilon$ in the matrix element, so that for $x<1$ the 
$d$-dimensional LO splitting function is identical to the 
four-dimensional one of \cite{am},
\begin{equation} \label{losplitd}
\Delta_T P_{qq}^{(0),d=4-2 \epsilon}(x) = \Delta_T P_{qq}^{(0)} (x) =
C_F \frac{2 x}{1-x} \hspace*{1cm} (x<1) \; .
\end{equation}
Note that this result for the $d$-dimensional LO splitting function is at
variance with the one of \cite{basim} in which residual terms $\sim \epsilon
(1-x)$ are present, presumably resulting from the assumption $k\cdot s=0$
in \cite{basim}. We will return to this point later. 

At NLO, the situation is more complicated. For the virtual graphs, 
besides terms involving $(k \cdot s)$ also powers of $(r \cdot s)$
appear, where $r$ is the loop momentum. To integrate these, one needs tensorial
two- and three-point functions. It turns out that all the functions we
need here already appeared in our unpolarized \cite{ev} and longitudinally
polarized \cite{wv} calculations, so that there is no really new integral
to be calculated. When calculating the squared matrix elements for
the {\em real} diagrams, one encounters the terms $(k \cdot s)^2$, 
$(k \cdot s)(l_1 \cdot s)$ and $(l_1 \cdot s)^2$, where $l_1$ is the momentum 
of one of the outgoing 'unobserved' particles. One can integrate these terms 
in two steps: first, one integrates over the terms $(l_1 \cdot s)$,
$(l_1 \cdot s)^2$. These integrations have lengthy expressions, given in 
detail in Appendix A. After this integration, the squared matrix element
still depends on $(k \cdot s)^2$ for which one finds, similarly to
(\ref{repl1}):
\begin{equation} \label{repl2}
(k \cdot s)^2 \rightarrow -\frac{1}{2 (1-\epsilon)} \left( k^2 -2 x 
(p\cdot k) \right) \; .
\end{equation}
After this substitution, the resulting expression for the matrix element 
squared contains just terms that are familiar from the unpolarized 
calculation. It can then be integrated using the techniques developed in 
\cite{ev,wv}.
\item For a transversely polarized {\em anti}quark with a spinor $v(p,s)$ 
one has expressions identical to (\ref{g51}),(\ref{g52}). Thus -- 
unlike the longitudinally polarized case \cite{wv} -- there is no extra 
minus sign when calculating the function $\Delta_T P_{q\bar{q}}^{(1)}$.
\item The endpoint contributions to the transversity splitting functions,
i.e. the contributions $\sim \delta (1-x)$, are necessarily the same as in
the unpolarized and longitudinally polarized cases since they are just
provided by the residue of the pole of the full quark propagator
\cite{cfp,ev,wv}.
\end{itemize}
We are now in a position to present the final results.
\section{Results}
For completeness, we begin with the full one-loop result \cite{am}:
\begin{equation}
\Delta_T P_{qq}^{(0)} (x) = C_F \left[ \frac{2x}{(1-x)_+}+\frac{3}{2}
\delta (1-x) \right] \; ,
\end{equation}
where the $+$-prescription is defined in the usual way:
\begin{equation}
\int_0^1 dz f(z) \left( g(z) \right)_+ 
\equiv \int_0^1 dz \left( f(z)-f(1) \right) g(z) \:\:\: . 
\end{equation}
To write down our final result for the $\Delta_T P_{qq,\pm}^{(1)}$ we
introduce 
\begin{eqnarray} 
\delta_T P_{qq}^{(0)} (x) &=& \frac{2x}{(1-x)_+} \; , \\
S_2(x) &=& \int_{\frac{x}{1+x}}^{\frac{1}{1+x}} \frac{dz}{z} 
\ln \big(\frac{1-z}{z}\big) \nonumber \\
&=& -2 {\rm Li}_2 (-x)-2 \ln x \ln (1+x)+\frac{1}{2} \ln^2 x-
\frac{\pi^2}{6} \; ,
\end{eqnarray}
where ${\rm Li}_2 (x)$ is the dilogarithm. We then have (cf. Eq.~(\ref{pm})) 
in the $\overline{{\mbox{\rm MS}}}$ scheme
\begin{equation}
\Delta_T P_{qq,\pm}^{(1)} (x) \equiv \Delta_T P_{qq}^{(1)} (x) \pm 
\Delta_T P_{q\bar{q}}^{(1)} (x) \; ,
\end{equation}
where\footnote{Needless to say that the $+$-prescription is not 
needed if the function multiplying the factor $1/(1-x)_+$ is vanishing 
at $x=1$. Neither is it to be taken into account in the function 
$\delta_T P_{qq}^{(0)}(-x)$.}
\begin{eqnarray} 
\Delta_T P_{qq}^{(1)} (x) &=& C_F^2 \Bigg[ 1-x - \left( \frac{3}{2} + 
2 \ln (1-x) \right) \ln x \; \; \delta_T P_{qq}^{(0)}(x) \nonumber \\
&&\hspace*{1cm} + \left. \left( \frac{3}{8} -\frac{\pi^2}{2} + 6\zeta (3) 
\right) \delta (1-x) \right] \nonumber \\
&+& \frac{1}{2} C_F N_C \left[ - (1-x) + \left( \frac{67}{9} + \frac{11}{3} 
\ln x + \ln^2 x - \frac{\pi^2}{3} \right) \delta_T P_{qq}^{(0)}(x) \right. 
\nonumber \\
&& \hspace*{1.6cm} + \left. \left( \frac{17}{12} + \frac{11 \pi^2}{9} -
6 \zeta (3) \right) \delta (1-x) \right] \nonumber \\
&+&\frac{2}{3} C_F T_f  \left[ \left( - \ln x -\frac{5}{3} \right) \delta_T 
P_{qq}^{(0)}(x) - \left( \frac{1}{4} + \frac{\pi^2}{3} \right) \delta 
(1-x) \right] \; , \label{ppp} \\
\Delta_T P_{q\bar{q}}^{(1)} (x) &=& C_F \left( C_F - \frac{1}{2} N_C \right)
\Bigg[ -(1 -x) + 2 S_2 (x) \delta_T P_{qq}^{(0)}(-x) \Bigg] \; , \label{ppm}
\end{eqnarray}
where $C_F=4/3$, $N_C=3$, $T_f=f T_R=f/2$ and $\zeta (3) \approx 
1.202057$. 

For checking Soffer's inequality via Eq.~(\ref{masterx}) we still need
the ${\cal O} (\alpha_s)$ corrections to the short-distance subprocess cross 
section for the transversely polarized Drell-Yan, $\Delta_T C_q^{{\small
{\rm DY}}} (x)$, in the $\overline{{\mbox{\rm MS}}}$ scheme.
The corresponding result was first presented in \cite{cont1} where,
however, dimensional {\em reduction} rather than dimensional regularization
was used. The translation of the result to dimensional regularization was
provided in \cite{basim}. As a check of the expression in \cite{basim}, we 
employ an earlier result for the transversely polarized Drell-Yan 
cross section at ${\cal O} (\alpha_s)$ which was obtained in \cite{alex}
by assuming off-shell gluons to regularize the appearing poles. Studying 
in detail the structure of the collinear singularities for both dimensional 
and off-shell regularization, it is straightforward to transform the result
from one regularization scheme to the other. Starting from the result
in \cite{alex}, we obtain in this way for the $\overline{{\mbox{\rm MS}}}$ 
scheme:
\begin{eqnarray} \label{cqdy}
\Delta_T C_q^{{\small {\rm DY}}} (x) &=& C_F \left[ 8x \left( 
\frac{\ln (1-x)}{1-x} \right)_+ - \frac{4 x \ln x}{1-x}-\frac{6 x 
\ln^2 x}{1-x} + 4 (1-x) \right. \nonumber \\
&& \hspace*{0.9cm} + \left( \frac{2}{3} \pi^2 -8 \right) \delta (1-x) 
\Bigg] \; .
\end{eqnarray}
This result indeed coincides with the one in \cite{basim} for the choice
$\Delta_T d= - \delta (1-x)$ in that paper. On the other hand, the 
expression for $\Delta_T d$ that is claimed in \cite{basim} to provide 
the link to the $\overline{{\mbox{\rm MS}}}$ scheme (in dimensional 
regularization) is $\Delta_T d=-\delta (1-x) + 2 (1-x)$. The reason for this 
difference lies in the discrepancy between our calculation and 
the one of \cite{basim} for the $(4-2\epsilon)$-dimensional LO splitting 
function (see the remark after Eq.~(\ref{losplitd})). We note that
the correctness of our result (\ref{losplitd}) for this quantity is 
corroborated by the observation that the $d$-dimensional $2\rightarrow 3$ 
matrix element squared for the process $\vec{q}\vec{\bar{q}} \rightarrow 
\mu^+ \mu^- g$ (with transversely polarized incoming (anti)quarks) nicely 
factorizes into the product of the $d$-dimensional $2\rightarrow 2$ matrix 
element squared for $\vec{q}\vec{\bar{q}} \rightarrow \mu^+ \mu^-$ times the 
splitting function $\Delta_T P_{qq}^{(0),d=4-2 \epsilon}$ of 
Eq.~(\ref{losplitd}), when the collinear limit of the gluon to be parallel
to one of the incoming quarks is taken properly. This clearly demonstrates
again the correctness of (\ref{losplitd}). The result of \cite{basim} 
for the transversely polarized Drell-Yan cross section at NLO (obtained 
within dimensional regularization) therefore necessarily corresponds to a 
different (non-$\overline{{\mbox{\rm MS}}}$) factorization scheme.

The remaining NLO ingredients for Eq.~(\ref{masterx}) are 
the unpolarized quantities $P_{qq}^{(1)} (x)$ and $C_q^{{\small {\rm DY}}} 
(x)$, which can be found in \cite{cfp,ev} and \cite{aem}, respectively.
Inserting everything into Eq.~(\ref{masterx}) it turns out that all 
distributions at $x=1$ ($+$-prescriptions and $\delta$-functions) drop out,
so that everything is regular at $x=1$. We can therefore plot the result 
for 
\begin{equation} \label{master1}
\tilde{D} (x) \equiv \left( P_{qq}^{(1)} (x)-\Delta_T P_{qq,-}^{(1)} (x) 
\right) - \frac{\beta_0}{4} \left( C_q^{{\small {\rm DY}}} (x) - 
\Delta_T C_q^{{\small {\rm DY}}} (x) \right) 
\end{equation}
directly in $x$-space. This is done in Fig.~1 for $f=3$ active flavours.
One can clearly see that $\tilde{D}(x)$ is always positive, which  
demonstrates the validity of Soffer's inequality for valence densities 
at the NLO level in the sense that if Soffer's inequality is valid at the
input scale, it will not be broken at any higher $Q^2$. As becomes visible 
from Fig.~1, $\tilde{D}(x)$ diverges at $x\rightarrow 0$. 
This feature is a consequence of a rather peculiar small-$x$ behaviour 
of the NLO transversity splitting functions and subprocess cross sections: 
the quantities in Eqs.~(\ref{ppp}),(\ref{ppm}),(\ref{cqdy})
all have a constant limiting behaviour at $x\rightarrow 0$. All terms
that rise logarithmically in $x$ are dampened by a factor $x$ from the 'LO' 
splitting term $\delta_T P_{qq}^{(0)}(x)$. This behaviour is in contrast to
the one for $P_{qq}^{(1)} (x)$, $C_q^{{\small {\rm DY}}} (x)$ in
(\ref{master1}), hence the rise of $\tilde{D}(x)$ at small $x$. 
More precisely, we find from (\ref{ppp}),(\ref{ppm}) for $x\rightarrow 0$
\begin{eqnarray} \label{smallx1}
\Delta_T P_{qq,-}^{(1)} (x) &\approx& C_F \left( 2 C_F-N_C \right) + 
2 C_F^2 x \ln^2 x \; , \\ \label{smallx2}
\Delta_T P_{qq,+}^{(1)} (x) &\approx& -2 C_F \left( C_F-N_C \right)
x \ln^2 x \; , 
\end{eqnarray}
that is, the constant terms even cancel in $\Delta_T P_{qq,+}^{(1)} (x)$.
We note that the logarithmic terms in (\ref{smallx1}),(\ref{smallx2}) can be 
recovered from Eq.~(3.15) of \cite{kmss} where a resummation of 
small-$x$ double-logarithms for the transversity densities was 
performed\footnote{Here one has to correct for a missing factor
$1/2$ in front of the term $f_V^-(\omega)$ in Eq.~(3.15) of \cite{kmss}
that got lost in the derivation of this equation from \cite{kold}.
I am thankful to A.\ Vogt for helpful communications on this point.}.
The constant terms in (\ref{smallx1}), however, were not anticipated to appear 
in ordinary Altarelli-Parisi evolution in \cite{kmss}.
The very mild small-$x$ behaviour of Eqs.~(\ref{smallx1}),(\ref{smallx2}) will 
lead to a strong suppression of the transversity distributions with respect 
to the unpolarized and longitudinally polarized ones during $Q^2$ evolution 
at small $x$.
\begin{figure}[htb]
\epsfig{file=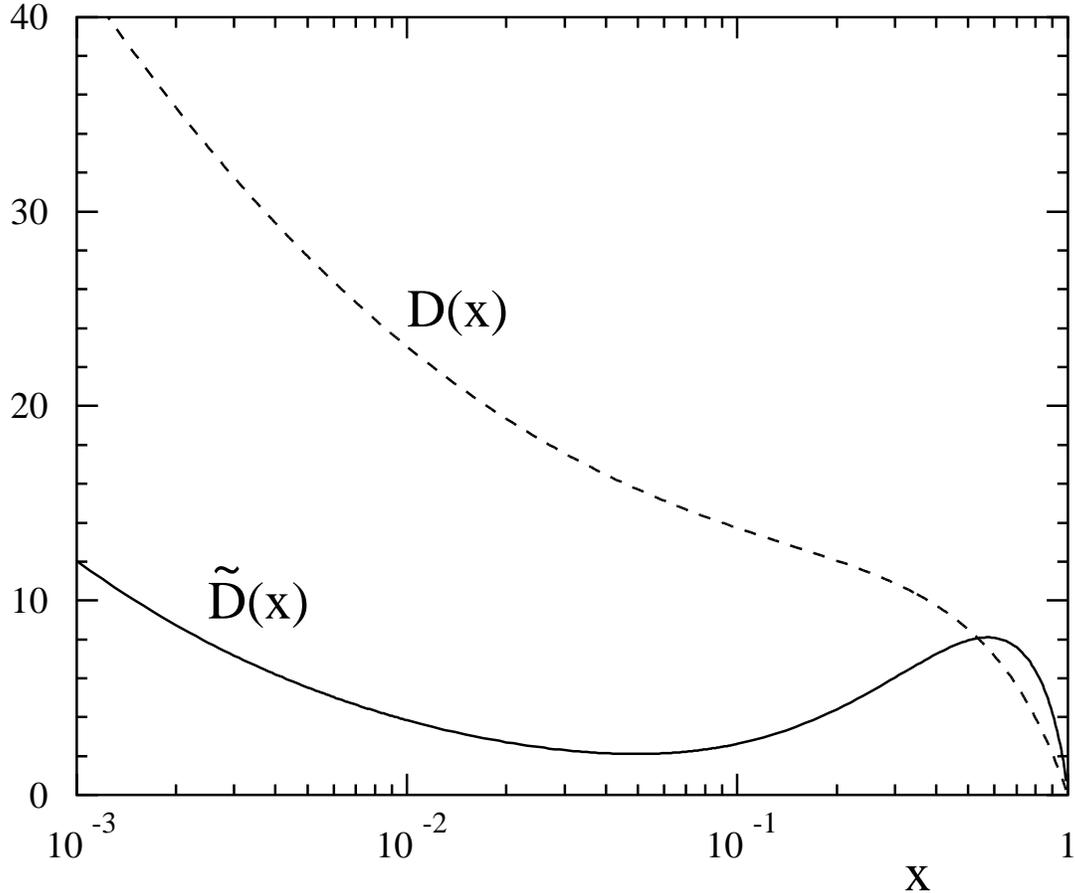,width=16cm}
\caption{The functions $\tilde{D}(x)$, $D (x)$ 
(see Eqs.~(\ref{master1}),(\ref{masterxms})) for $f=3$ active flavours.}
\end{figure}

As discussed in Sec.~2 we have chosen a factorization scheme in which 
the NLO Drell-Yan cross sections for the unpolarized, the longitudinally
polarized and the transversely polarized cases all retain their respective
LO forms. We have done this to deal with quantities that are manifestly
scheme-independent. From a practical point of view, this may not be 
very useful since the unpolarized and longitudinally polarized parton
densities are usually defined in the $\overline{{\mbox{\rm MS}}}$ scheme.
In particular, when building models for the transversity densities, one
would like to know whether Soffer's inequality is also preserved under
NLO evolution in the $\overline{{\mbox{\rm MS}}}$ scheme. If this is 
the case then it appears likely to make sense to assume validity of Soffer's 
inequality also for the $\overline{{\mbox{\rm MS}}}$ NLO input parton 
densities and not only for the parton densities defined in our scheme. 
For this reason the dashed line in Fig.~1 shows only the quantity
\begin{equation} \label{masterxms}
D (x) \equiv P_{qq}^{(1)} (x)-\Delta_T P_{qq,-}^{(1)} (x) \; ,
\end{equation} 
which corresponds to performing the evolution in the $\overline{{\mbox{\rm 
MS}}}$ scheme. As can be seen, Soffer's inequality will also be maintained 
here. 

For convenience, we present the Mellin-$n$ moments of the transversity
NLO quantities $\Delta_T P_{qq,\pm}^{(1)}$, $\Delta_T 
C_q^{{\small {\rm DY}}}$ in Appendix B. These are useful for a numerical
evaluation of our results. Fig.~2 shows the moments of the splitting 
functions versus real $n \geq 1$. As can be seen, the moments are all
negative. Furthermore, the numerical contribution of the splitting function
$\Delta_T P_{q\bar{q}}^{(1),n}$ is very small for $n\geq 1$. For comparison 
we also show the moments of the LO transversity splitting function. The 
expression for the first ($n=1$) moment of $\Delta_T P_{qq,-}^{(1)}$ turns 
out to be 
\begin{equation}
\Delta_T P_{qq,-}^{(1),1} = \frac{19}{8} C_F^2 - \frac{257}{72} C_F N_C
+\frac{13}{18} C_F T_f \; .
\end{equation}
This quantity participates in the NLO evolution of the nucleon's tensor 
charge which is given by the first moment of the sum of the valence 
transversity densities \cite{jaffe}.
\begin{figure}[htb]
\epsfig{file=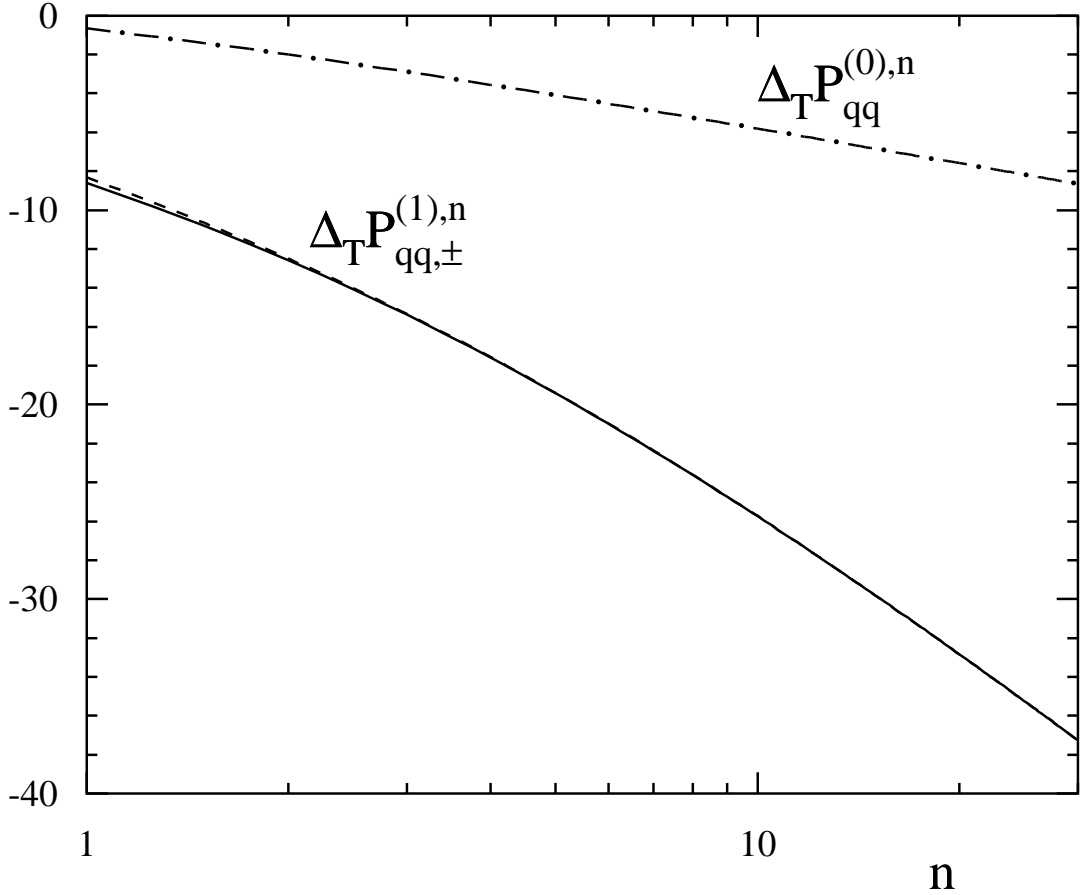,width=16cm}
\caption{The Mellin-$n$ moments of the NLO transversity splitting functions
$\Delta_T P_{qq,-}^{(1)}$ (solid line) and $\Delta_T P_{qq,+}^{(1)}$
(dashed line) versus real $n\geq 1$. For comparison the dash-dotted curve 
shows the LO result $\Delta_T P_{qq}^{(0),n}$.}
\end{figure}
\section{Conclusions}
We have presented a calculation of the next-to-leading order splitting 
functions $\Delta_T P_{qq,\pm}^{(1)}(x)$ for the $Q^2$ evolution of 
transversity parton densities. The method we have used here is the one of
\cite{egmpr,cfp} which was applied to the unpolarized and longitudinally
polarized cases in \cite{cfp,fp} and \cite{wv}, respectively.
We have used our results to examine the preservation of Soffer's inequality
for valence quark densities under NLO $Q^2$ evolution. It turned out that
(in a suitably defined factorization scheme)
the inequality is indeed maintained at any higher $Q^2$ if it is satisfied 
at the input scale, which provides strong support for its correctness.
In other words, ${\cal O} (\alpha_s)$ corrections do not seem to invalidate 
the inequality. We note again that extending 
our results to non-valence quark densities affords inclusion of the full
singlet evolution of the unpolarized and longitudinally polarized densities
and thus has to be subject to a detailed numerical study. Combining our 
finding for the valence part with previous experience from LO \cite{bar} 
indicates that no surprises concerning Soffer's inequality are expected in the 
singlet sector.
\section*{Note added}
After completing this work we received the papers \cite{new1,new2} in which 
the two-loop anomalous dimensions for the transversity distributions were
calculated using the Operator Product Expansion. The obtained results should 
correspond to the NLO splitting functions we have calculated. Indeed, 
our results in Mellin-$n$ space in Eq.~(B.1) are exactly identical to
those in \cite{new2} and \cite{new1} (after revision of that paper).
We note that our results were first presented on the `Ringberg Workshop on 
High Energy Polarization Phenomena', Schlo{\ss} Ringberg, Germany,
25-28 February 1997.
\section*{Acknowledgements}
I am grateful to J.C.\ Collins, X.\ Ji, J.\ Soffer and M.\ Stratmann 
for helpful discussions and to B.\ Kamal for useful correspondence on his 
calculation of the NLO corrections to the transversely polarized Drell-Yan 
cross section.
\section*{Appendix A}
\appendix
\setcounter{equation}{0}
\renewcommand{\theequation}{A.\arabic{equation}}
In this appendix we provide the expressions for the integrations over
the terms $(l_1 \cdot s)$ and $(l_1 \cdot s)^2$. Defining 
\begin{equation}
\lambda \equiv 2 \left( -(1-x) (l_1 \cdot p)+\frac{k^2}{2} - (p\cdot k)
\Bigg( 1-\frac{l_1 \cdot n}{pn} \Bigg) \right) \; , 
\end{equation}
we find:
\begin{eqnarray}
(l_1 \cdot s) &\rightarrow& -\frac{1}{2} \; \; 
\frac{\lambda}{k^2-2 x (p\cdot k)} (k \cdot s) \; , \\
(l_1 \cdot s)^2 &\rightarrow& \frac{1}{1-2 \epsilon} \left( 2 (l_1 \cdot p)
\frac{l_1 \cdot n}{pn} + \frac{\lambda^2}{4 (k^2-2 x (p\cdot k))}\right)
\left[ 1+ \frac{(k\cdot s)^2}{k^2-2 x (p\cdot k)} \right] \nonumber \\
&&+\frac{\lambda^2 \; (k\cdot s)^2}{4 (k^2-2 x (p\cdot k))^2} 
\end{eqnarray}
as the effective substitutions in the squared matrix element. For kinematics
and the notation of the momenta see \cite{wv}.
\section*{Appendix B}
\appendix
\setcounter{equation}{0}
\renewcommand{\theequation}{B.\arabic{equation}}
In this appendix we present the Mellin-$n$ moments of the NLO quantities
$\Delta_T P_{qq,\pm}^{(1)} (x)$, $\Delta_T C_q^{{\small {\rm DY}}} (x)$
(in the $\overline{{\mbox{\rm MS}}}$ scheme) which are useful for a numerical 
evaluation of our results:
\begin{eqnarray}
\Delta_T P_{qq,\eta}^{(1),n} &=& C_F^2 \left[ \frac{3}{8} + 
\frac{1-\eta}{n (n+1)} - 3 S_2 (n) - 4 S_1 (n) \left( S_2 (n) - 
S'_2 (\frac{n}{2}) \right) - 8 \tilde{S}(n) + S'_3 (\frac{n}{2}) \right]
\nonumber \\
&+& \frac{1}{2} C_F N_C \left[ \frac{17}{12} - \frac{1-\eta}{n (n+1)}
- \frac{134}{9} S_1 (n) + \frac{22}{3} S_2 (n) \right. \nonumber \\
&&\left. + 4 S_1 (n) \left( 2 S_2 (n) - S'_2 (\frac{n}{2}) \right)
+ 8 \tilde{S}(n) - S'_3 (\frac{n}{2}) \right] \nonumber \\
&+& \frac{2}{3} C_F T_f \left[ - \frac{1}{4} + \frac{10}{3} S_1 (n) - 
2 S_2 (n) \right] \; , \\
\Delta_T C_q^{{\small {\rm DY}},n} &=& C_F \left[ \frac{4}{n(n+1)}
+ 4 S_1^2 (n) + 12 \left( S_3 (n) - \zeta (3) \right) - 8 + 
\frac{4}{3} \pi^2 \right] \; ,
\end{eqnarray} 
where $\eta\equiv\pm$. The sums appearing here are defined by
\begin{eqnarray}
S_k (n) &\equiv& \sum_{j=1}^n \frac{1}{j^k} \nonumber \; , \\
S'_k (\frac{n}{2}) &\equiv& 2^{k-1} \sum_{j=1}^n \frac{1+(-1)^j}{j^k}
\nonumber \; \\
\tilde{S} (n) &\equiv& \sum_{j=1}^n \frac{(-1)^j}{j^2} S_1 (j) \; .
\end{eqnarray}
Their analytic continuations to arbitrary Mellin-$n$ (which depend on 
$\eta$) can be found in \cite{grv}.

\end{document}